\documentclass[11pt]{article}
\usepackage{amsmath, amssymb, amsthm}
\usepackage{graphicx}
\usepackage{natbib}
\usepackage{bm}
\usepackage{booktabs}
\usepackage{algorithm}
\usepackage{algorithmic}
\usepackage{authblk}
\usepackage{xcolor}
\usepackage{appendix}

\usepackage{listings}
\begin{document}

{
  \title{\bf On the Computation of Normalized Power Priors}
  \author[1]{Fang Chen\thanks{Corresponding author: fangk.chen@sas.com}} 
  \author[2]{Kaifeng Lu}
  \affil[1]{SAS Institute Inc.}
  \affil[2]{BeOne Medicines}
  \date{} 
  \maketitle
} 

\begin{abstract}

The normalized power prior provides a principled framework for incorporating 
historical data into Bayesian inference while preserving coherence, but 
its routine application has been hindered by the need to evaluate an 
intractable normalizing constant function $C(a_{0})$. Existing approaches typically 
rely on model-specific marginal likelihood calculations, numerical 
integration over grids, or auxiliary sampling schemes implemented outside 
standard Bayesian software. In this paper, we present a computational 
perspective that exploits a simple and underutilized functional identity: 
under the unnormalized power prior, the marginal distribution of the power 
parameter is proportional to the normalizing constant of the normalized 
power prior. Leveraging this relationship, we propose a sampling-based 
strategy to approximate the normalized power prior using output from 
generic Markov chain Monte Carlo (MCMC) algorithms. The resulting approximation 
can be implemented entirely within all general Bayesian software 
packages (such as PROC MCMC, BUGS, JAGS, Stan, or NIMBLE), without 
requiring explicit marginal likelihood evaluation. 
Several illustrative examples demonstrate the practicality of the 
approach and highlight its potential to facilitate the routine use of 
normalized power priors in applied settings.

\end{abstract}

\section{Introduction}

The increasing availability of historical clinical, observational, and
real-world data has intensified interest in methods that formally and
flexibly incorporate such information into modern Bayesian
analyses. When used appropriately, historical data can sharpen
inference, reduce uncertainty, improve trial efficiency, and, in some
settings, reduce the required sample size of a prospective
study. However, inappropriate borrowing may distort inference or lead to
misleading conclusions when historical and current data
differ. Consequently, methodological frameworks that (i) allow the
analyst to control the degree of borrowing and (ii) quantify uncertainty
in that borrowing have become essential tools in contemporary
biostatistics. 

The "power prior", originally introduced by \citet{IbrahimChen2000},
provides one of the most influential approaches for dynamic historical
borrowing. Given historical data $D_0$ and current data $D$, the power
prior augments the likelihood for the current study with a down-weighted
contribution from historical observations. The amount of borrowing is
governed by a parameter $a_0 \in [0,1]$, where $a_0 = 0$ corresponds to
discarding historical information entirely and $a_0 = 1$ corresponds to
full pooling. This construction offers a transparent and interpretable
means of discounting historical evidence based on its relevance,
similarity, or quality. 

In practice, analysts frequently fix $a_0$ at a prespecified value
(e.g., 0.3 or 0.5), resulting in the original power prior (OPP). This
formulation is straightforward to implement in any general Bayesian
software: you combine the historical and current data sets into a larger
data set and fit a regular Bayesian model, albeit with historical
log-likelihoods multiplied by $a_0$. By contrast, allowing $a_0$ to be
random enables a data-driven assessment of how much borrowing is
supported, leading to richer inference and posterior adaptivity. Yet
this extension creates a fundamental challenge. The fully Bayesian
treatment requires normalizing the joint prior so that the prior for
$a_0$ integrates correctly, and this introduces a normalizing function,
\[
C(a_0) = \int L(\theta | D_0)^{a_0} \, \pi_0(\theta)\, d\theta,
\]
which depends on $a_0$ and is typically analytically intractable. The
resulting formulation, termed the normalized power prior (NPP)
\citep{Duan2006, Neuenschwander2009}, is theoretically principled but
practically difficult because $C(a_0)$ must be computed for every 
proposed value of $a_0$ during Markov chain Monte Carlo (MCMC) sampling. 

A variety of numerical strategies have been proposed for evaluating or
approximating $C(a_0)$, including Laplace approximations, importance
sampling, and bridge sampling \citep{CarvalhoIbrahim2021}. While
effective in controlled examples, these methods often require
specialized coding, repeated fitting of historical-data models, or
external precomputation of $C(a_0)$ over a grid of $a_0$
values. Consequently, none of these approaches integrate seamlessly into
general-purpose Bayesian engines such as SAS PROC MCMC, JAGS, BUGS,
Stan, or NIMBLE. This gap between theoretical appeal and practical
feasibility has limited the routine use of NPP in applied settings. 

The objective of this manuscript is to bridge this gap by developing a
computational framework for implementing the normalized power prior
directly inside standard Bayesian modeling software, without requiring
external marginal-likelihood estimation. We propose a sampling-based
method that exploits a key functional identity: under a uniformly distributed
$a_0$, $C(a_0)$ is functionally proportional to the marginal distribution of $a_0$ in 
the unnormalized power prior (UnPP) using the historical data. By "functional", we 
mean that this equivalency is true when the unnormalized power prior is viewed as a pseudo 
likelihood function that enables inference on $a_0$ given $D_0$. 
This allows us to estimate $C(a_0)$ using MCMC samples from an
easily implemented UnPP model. We then fit flexible parametric
mixture models to approximate $C(a_0)$, yielding a plug-in function
$\tilde C(a_0)$ that can be used within a fully Bayesian NPP
analysis. Importantly, this approach requires only simple modifications
to existing power-prior models and is compatible with any environment
allowing user-defined log-likelihoods. In routine settings, the added
work is modest: estimate a mixture for $C(a_0)$ (for example, with PROC
FMM) and insert $\log \tilde C(a_0)$ into an existing OPP program. In
more difficult cases, additional approximation steps are needed, but
they remain feasible in general Bayesian software.

We show that, in ideal situations, the proposed method produces accurate
posterior inference for both $a_0$ and model parameters; in scenarios
where $C(a_0)$ decays rapidly or exhibits strong convexity, we propose simple approximation methods. Our findings show that, when
implemented with refinements, the sampling-based approach provides
accuracy comparable to more computationally intensive methods while
retaining simplicity and portability. 

This manuscript is organized as follows. Section~\ref{sec:theory}
provides a review of the power prior framework, including the
normalized formulation and computational obstacles.
Section~\ref{sec:methods} demonstrates and examines how the proposed
method performs under different scenarios. 
We conclude with a discussion of practical considerations,
software integration, and opportunities for future extensions 
in Section~\ref{sec:discussion}. The Appendix
includes SAS code and computational details for implementing the method
in practice.

\section{Background}
\label{sec:theory}

\subsection{Various Types of Power Priors}

The power prior framework offers a principled Bayesian mechanism for
synthesizing historical and current information while controlling the
degree of borrowing through a tuning parameter $a_0 \in [0,1]$. In this
section, we review the foundational components of the original power
prior (OPP), the unnormalized power prior (UnPP), and the normalized
power prior (NPP).

Let $D_0$ represent historical data with likelihood $L(\theta|D_0)$, 
and let $\pi_0(\theta)$ denote an initial prior for the model
parameters $\theta$. The original power prior, introduced by
\citet{IbrahimChen2000}, defines a new prior distribution as 
\[
\pi(\theta | D_0, a_0) \propto L(\theta | D_0)^{a_0} \pi_0(\theta),
\qquad a_0 \in [0,1].
\]

This construction down-weights the contribution of the historical
likelihood by exponentiating it with $a_0$.  When $a_0 = 0$, historical
information is completely ignored; when $a_0 = 1$, the prior corresponds
to full pooling, effectively treating $D_0$ as if it were part of the
current dataset.

Under current data $D$, the posterior distribution becomes
\[
\pi(\theta | D, D_0, a_0)
\propto
L(\theta | D)\, L(\theta | D_0)^{a_0}\, \pi_0(\theta),
\]
which is straightforward to implement in any Bayesian software: you
combine $D$ and $D_0$ into a larger data set, and this becomes a regular
Bayesian analysis with prior $\pi_0(\theta)$ and observations from $D_0$
weighted by $a_0$. 

However, the OPP requires $a_0$ to be fixed. Choosing $a_0$ arbitrarily
introduces subjectivity, and treating $a_0$ as fixed does not capture
uncertainty about the relevance of historical information.

To overcome the limitation of a fixed $a_0$, one may assign a prior
distribution $\pi_A(a_0)$ over $a_0$ and jointly estimate $(\theta,
a_0)$. The resulting joint prior is
\begin{equation}
\label{eq:unpp-prior}
\tilde{\pi}(\theta, a_0 | D_0)
\propto
L(\theta | D_0)^{a_0}
\pi_0(\theta)
\pi_A(a_0).
\end{equation}
This formulation is called the unnormalized power prior (UnPP).

Although the UnPP is simple to implement, it is theoretically
problematic. The induced marginal prior for $a_0$ is distorted, and the
resulting posterior may not correspond to a proper Bayesian updating
scheme. In particular, the distribution of $a_0$ often becomes
artificially concentrated near the boundaries of $[0,1]$, leading to
biased borrowing.

\citet{Duan2006} and \citet{Neuenschwander2009} noted that the correct power prior with $a_0$ being 
random requires an additional normalizing constant function:
\begin{equation}
\label{eq:npp-prior}
\pi(\theta, a_0 | D_0)
=
\frac{L(\theta | D_0)^{a_0}}{C(a_0)}\,
\pi_0(\theta)\,
\pi_A(a_0),
\end{equation}
where $C(a_0)$ is defined as
\begin{equation}
\label{eq:npp-c}
C(a_0)
=
\int L(\theta | D_0)^{a_0} \pi_0(\theta) d\theta,
\end{equation}
ensuring that the conditional prior for $\theta$ given $a_0$ is properly
normalized:
\[
\int \pi(\theta | a_0, D_0)\, d\theta = 1.
\]

Under NPP, inference for $a_0$ is now properly Bayesian, with prior
$\pi_A(a_0)$ accurately reflected in the posterior. The posterior
distribution for $(\theta, a_0)$ given current data is
\begin{equation}
\label{eq:npp}
\pi(\theta, a_0 | D, D_0)
\propto
L(\theta | D)
\frac{L(\theta | D_0)^{a_0}}{C(a_0)}
\pi_0(\theta)
\pi_A(a_0).
\end{equation}

The primary barrier to routine use of NPP is that $C(a_0)$ must be
evaluated (or approximated) for many candidate values during MCMC
sampling. Direct computation is infeasible in all but the simplest
cases, as it involves a high-dimensional integral.

Existing approximations mostly use numerical methods to approximate 
the function. \citet{CarvalhoIbrahim2021} proposed to use importance 
and bridge sampling to evaluate the normalizing function over 
gridded values and use that as a dictionary in NPP analysis. 

However, these methods may be computationally expensive, sensitive to
tuning parameters, or incompatible with off-the-shelf Bayesian software
such as PROC MCMC, BUGS, JAGS, or Stan.

\subsection{Key Identity Underlying the Proposed Method}

We want to use sampling methods to estimate $C(a_0)$, which has the form of a marginal distribution of $a_0$ given $D_0$. However, in an UnPP, the parameter $a_0$ is independent of $D_0$ and one is technically unable to infer $a_0$ given $D_0$ (this can be seen by taking the integral of $\pi(\theta, a_0 | D_0)$ in Eq(\ref{eq:npp-prior}) with respect to $\theta$  and the $C(a_0)$ terms cancel each other out, leaving only $\pi_A(a_0)$).

A central insight of this manuscript is to utilize the fact that $C(a_0)$ is computationally equivalent to the marginal distribution of $a_0$ in an UnPP when $L(\theta|D_0)^{a_0}$ is treated as a psuedo likelihood function of $D_0$, with $\theta$ and $a_0$ as model parameters. 

From a computational perspective, if we treat the powered likelihood function $L(\theta|D_0)^{a_0}$ as a psuedo likelihood function, we can infer $a_0$ from $D_0$. In other words, 
\begin{eqnarray*}
\tilde{\pi}(a_0|D_0) &=& \int \tilde{\pi}(\theta, a_0 | D_0) d\theta \\
& \propto & \int 
L(\theta | D_0)^{a_0}
\pi_0(\theta)
\pi_A(a_0) d\theta, 
\end{eqnarray*}
where $\tilde{\pi}$ is the marginal posterior of the \emph{unnormalized} power prior when the data is assumed with this pseudo likelihood function. Under a uniform prior on $a_0$, this becomes proportional to the normalizing function: 
\[
\tilde{\pi}(a_0 | D_0) \propto C(a_0).
\]

This means that we can obtain samples, from the UnPP, whose distribution is proportional to $C(a_0)$. We then approximate the marginal parametrically and use the functional form in the NPP analysis.

\section{Methods}
\label{sec:methods}

This section uses simulations to demonstrate how the proposed method works. 
We demonstrate that the sampling-based approximation $\tilde C(a_0)$
reproduces the borrowing behavior induced by the true $C(a_0)$. We also 
use a regression example to discuss situations where sampling-based approaches
are not ideal and how to make improvements. 

\subsection{Binomial Example}

We illustrate the behavior of the normalized power prior using a simple
binomial example, for which the normalizing constant $C(a_0)$ is available
in closed form.

Let the historical data consist of $y_0$ successes out of $n_0$ trials:
\[
Y_0 | p \sim \mathrm{Binomial}(n_0,p),
\]
with a $\mathrm{Beta}(c,d)$ prior on $p$:
\[
p \sim \mathrm{Beta}(c,d).
\]

The power weighted historical likelihood is
\[
L(p | D_0)^{a_0}
\propto
\binom{n_0}{y_0}^{a_0}
p^{a_0 y_0}
(1-p)^{a_0(n_0-y_0)}.
\]

The normalizing function of the normalized power prior is therefore
\[
C(a_0)
=
\int_0^1
L(p | D_0)^{a_0}
\pi(p)\,dp,
\]
which has the closed form expression
\[
C(a_0)
=
\binom{n_0}{y_0}^{a_0}
\frac{B(a_0 y_0 + c,\; a_0(n_0-y_0) + d)}{B(c,d)},
\]
where $B(\cdot,\cdot)$ denotes the Beta function.

Taking logarithms yields
\[
\log C(a_0)
=
a_0 \log\binom{n_0}{y_0}
+ \log B(a_0 y_0 + c,\; a_0(n_0-y_0) + d)
- \log B(c,d).
\]

\begin{figure}[htbp]
  \centering
  \includegraphics[width=0.48\textwidth]{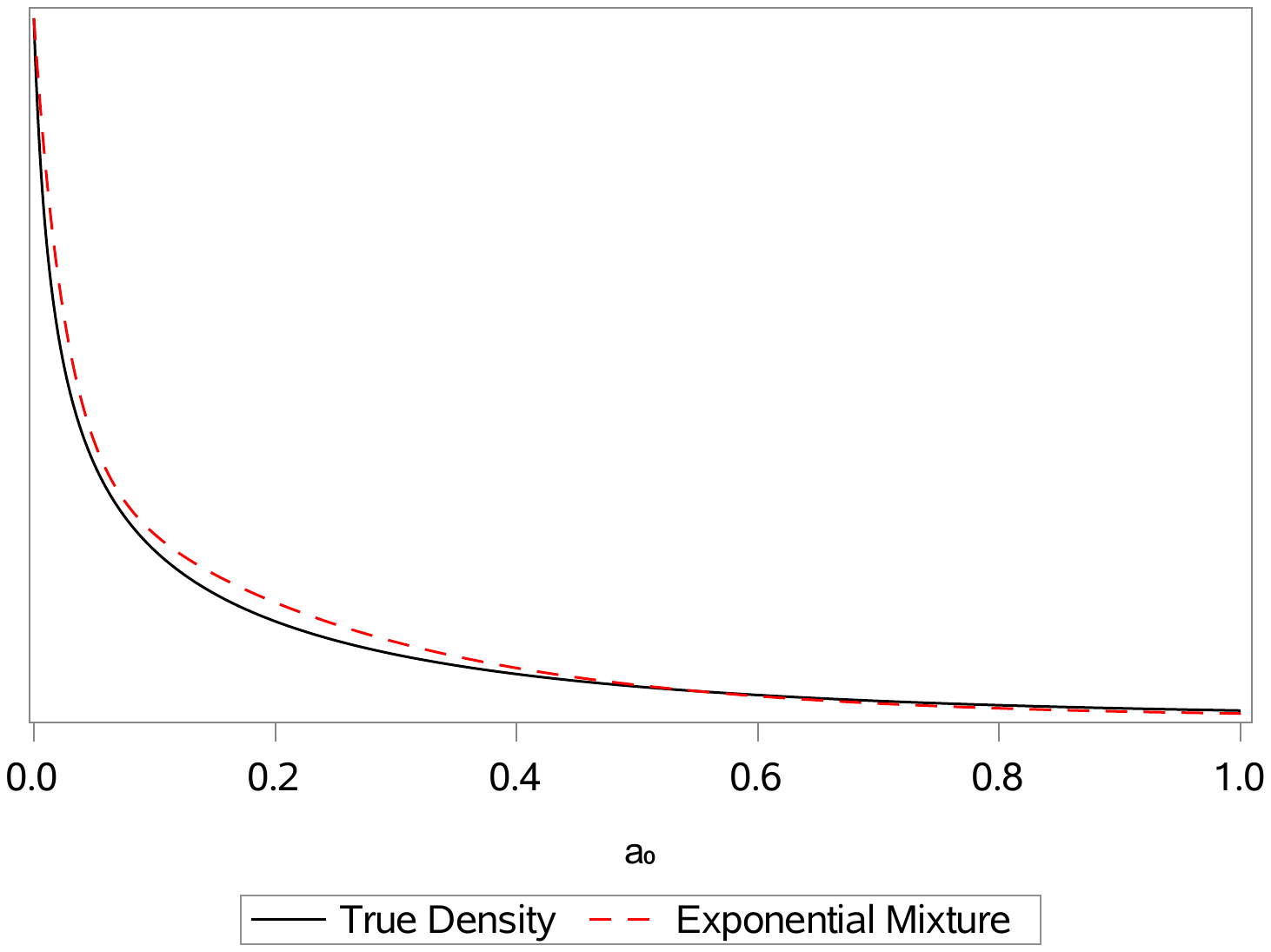}
  \includegraphics[width=0.48\textwidth]{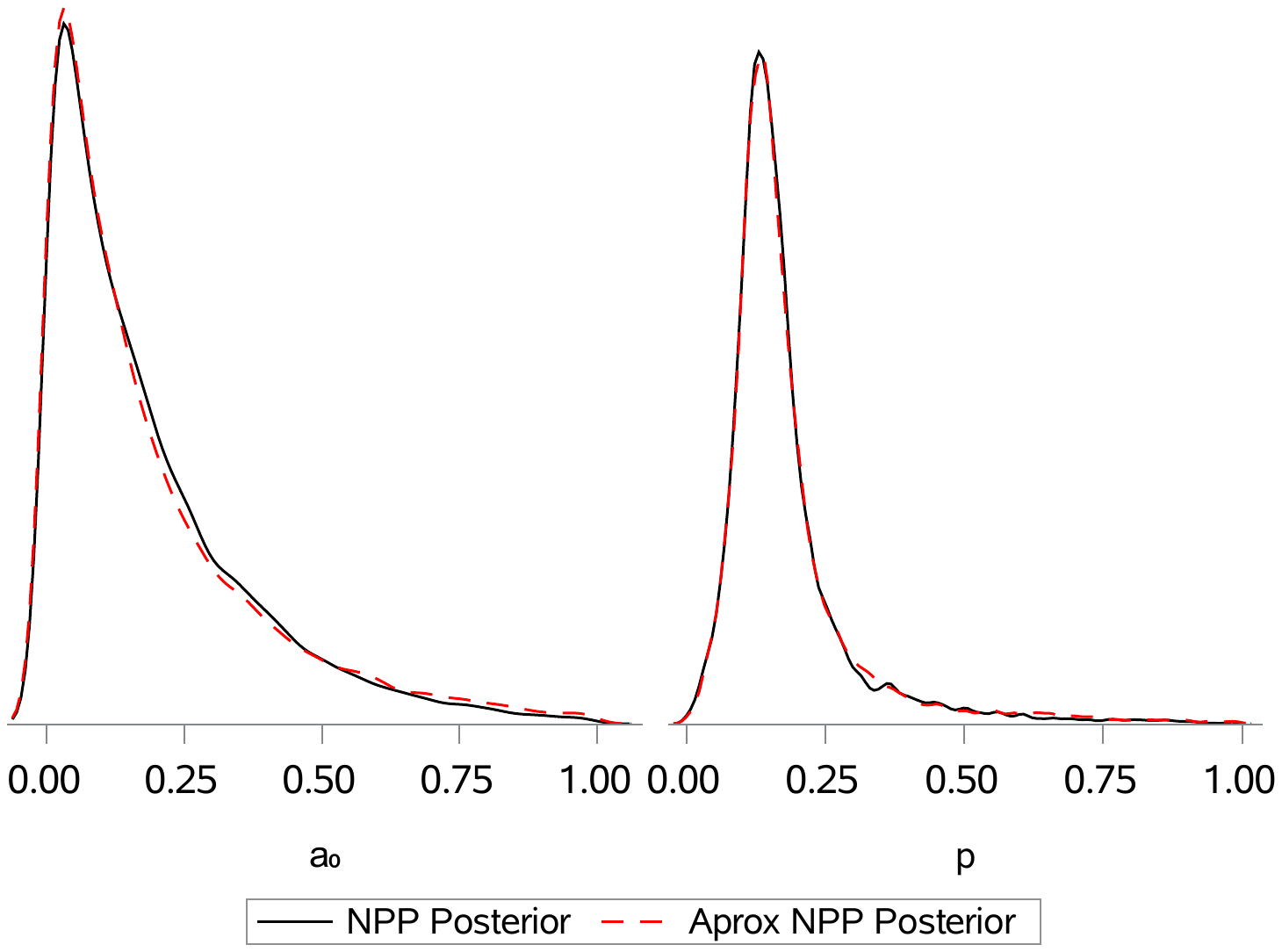}
  \caption{Left: Comparing the true (solid) and approximating (dashed) 
  normalizing constant function $C(a_0)$ in the Binomial example. 
  Right: Posterior distribution under the true NPP
  (solid) and its mixture-based approximation (dashed). The
  excellent agreement shows that the sampling-based approximation
  preserves the inferential quality of the NPP.}
  \label{fig:bernDist}
\end{figure}

The left graph in Figure~\ref{fig:bernDist} displays the true $C(a_0)$ 
(solid line) in a binomial setup together with a mixture approximation 
from the UnPP MCMC samples. The approximation is close. 
The analysis is based on a simple  binomial case, with $n_0 = 100$ 
and $y_0 = 20$ in $D_0$, and $n=200$ and $y = 20$ in $D$. 

The right graph in Figure~\ref{fig:bernDist} compares the posterior 
distributions of $a_0$ and $p$ using the true $C(a_0)$ versus the 
mixture-based $\tilde{C}(a_0)$ approximation. The close agreement 
between the two densities illustrates that, as expected in this 
setting, the mixture approximation closely captures the joint 
posterior of $a_0$ and $p$.

\subsection{Challenges in the Regression Case: Highly Skewed $a_0$ Posterior}

While in theory, a sampling-based approximation to $C(a_0)$ should be
sufficient because it provides an accurate approximation to the true
normalizing function. In practice, however, $C(a_0)$ is often observed
to decay extremely quickly. This leads to computational difficulties.

We consider the linear regression model with conjugate priors:
\begin{align*}
\sigma^2 &\sim \mathrm{IG}(\alpha_0,\gamma_0),\\[4pt]
\varepsilon_i | \sigma^2 &\sim \mathcal{N}(0,\sigma^2),\\[4pt]
\beta | \sigma^2 &\sim \mathcal{N}_p(\mu_0,\;\sigma^2 \Lambda_0),\\[4pt]
Y_i &= X_i^\top \beta + \varepsilon_i ,
\end{align*}
where $\Lambda_0$ is a positive-definite prior covariance matrix and
$N_0$ is the size of the historical dataset.

The priors have densities
\begin{align*}
\pi(\sigma^2)
&=
\frac{\gamma_0^{\alpha_0}}{\Gamma(\alpha_0)} (\sigma^2)^{-(\alpha_0+1)}
\exp\!\Big( -\frac{\gamma_0}{\sigma^2} \Big),
\\[6pt]
\pi(\beta|\sigma^2)
&=
(2\pi\sigma^2)^{-p/2}
|\Lambda_0|^{-1/2}
\exp\!\Big(
-\frac{1}{2\sigma^2}
(\beta-\mu_0)^\top\Lambda_0^{-1}(\beta-\mu_0)
\Big).
\end{align*}

The normalizing function of the normalized power prior for linear
regression has the following closed form:

\[
C(a_0)
= \frac{{\gamma_0}^{\alpha_0}}{\Gamma(\alpha_0)} |\Lambda_0^{-1}|^{1/2}
\frac{\Gamma(\alpha_n)}{\gamma_n^{\alpha_n}}
\,|\Lambda_n|^{1/2}
(2\pi)^{-a_0 N_0/2},
\]
where

\begin{align*}
\alpha_n &= \alpha_0 + \frac{1}{2}N_0 a_0 \\[4pt]
\Lambda_n^{-1} &= \Lambda_0^{-1} + X_\star^\top X_\star\\[4pt]
\gamma_n &= \gamma_0 + \frac{1}{2}\Bigl( Y_\star^\top Y_\star +
\mu_0^\top\Lambda_0^{-1}\mu_0 - \mu_n^\top \Lambda_n^{-1} \mu_n \Bigr) \\[4pt]
\mu_n &= \Lambda_n \bigl(\Lambda_0^{-1}\mu_0 + X_\star^\top
Y_\star\bigr) \\[4pt]
Y_\star &= \sqrt{a_0}\, Y_0 \\[4pt]
X_\star &= \sqrt{a_0}\, X_0 
\end{align*}

\begin{figure}[htbp]
  \centering
  \includegraphics[width=4in]{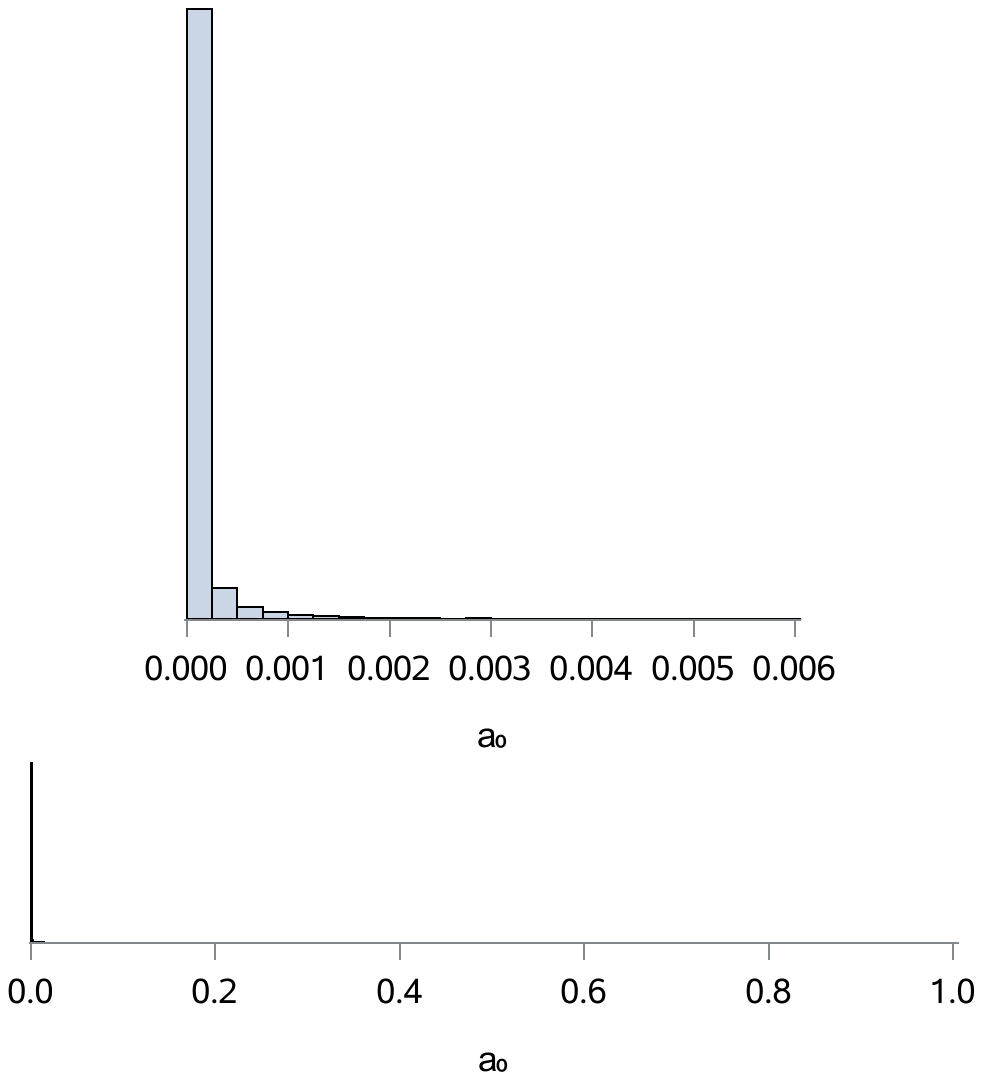}
  \caption{Unnormalized power prior posterior for $a_0$ in a regression
  example. The extreme skewness illustrates the difficulty of capturing
  the global structure of $C(a_0)$ using sampling-based methods.}
  \label{fig:regCa0}
\end{figure}

Figure~\ref{fig:regCa0} shows a representative posterior for $a_0$ under
the UnPP model for a regression dataset. The distribution is sharply
concentrated near zero, which becomes particularly challenging for
numerical estimation of $C(a_0)$, since only a narrow region of the
parameter space is meaningfully explored by MCMC.

\begin{figure}[htbp]
  \includegraphics[width=0.48\textwidth]{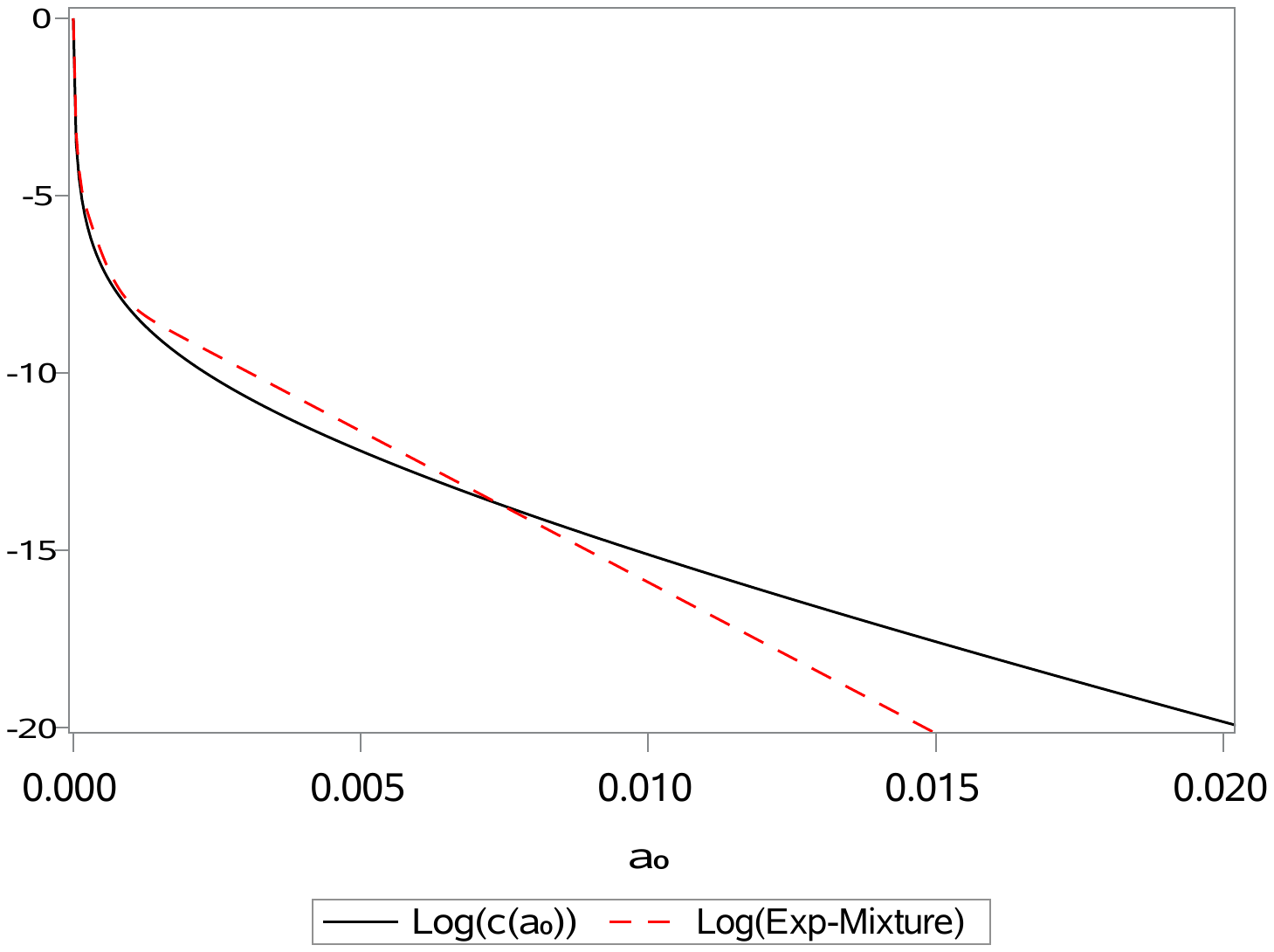}
  \hfill
  \includegraphics[width=0.48\textwidth]{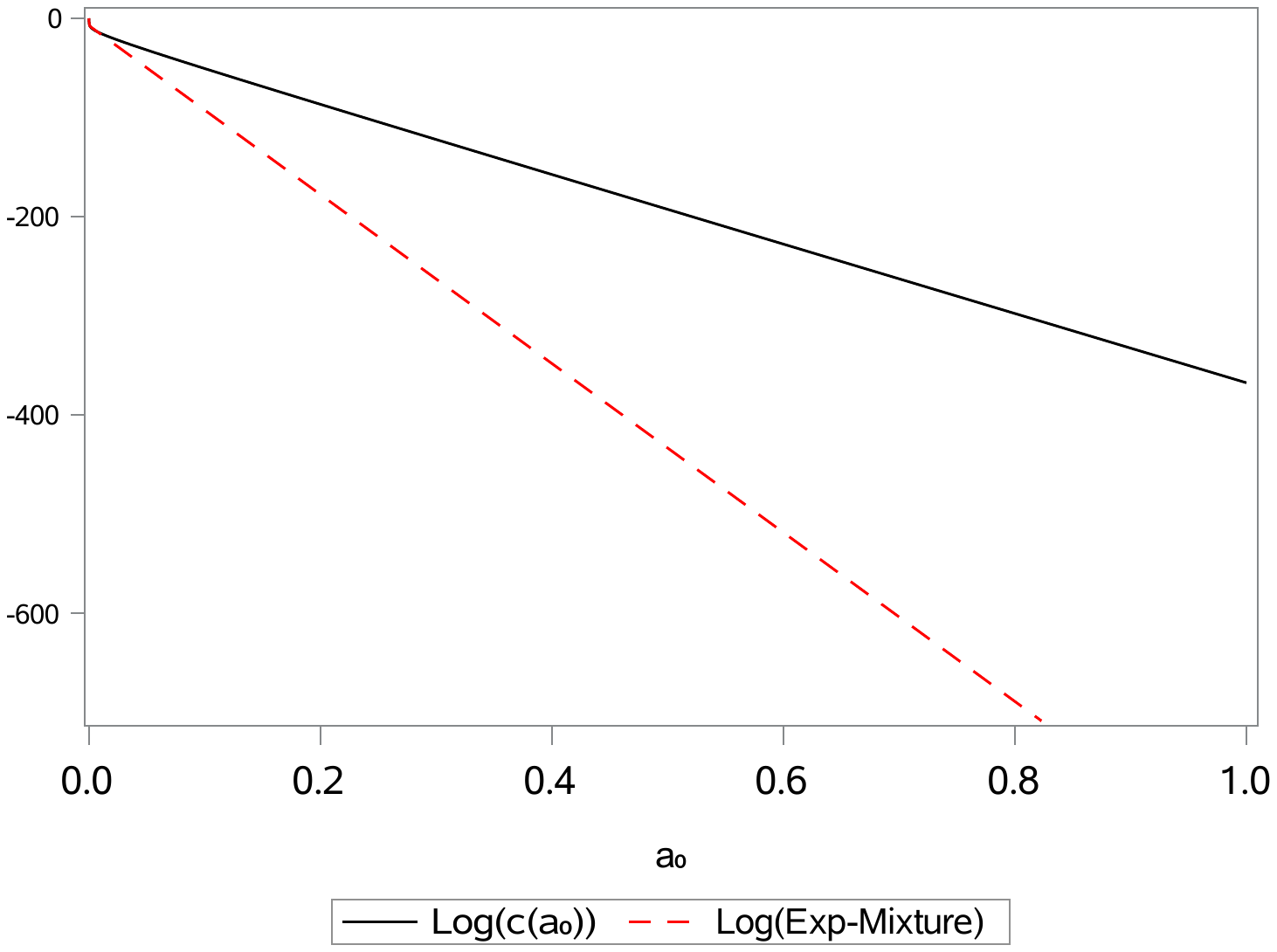}
  \caption{Left: close-up comparison between true and approximate $\log
  C(a_0)$ near $a_0 = 0$. Right: global comparison across the full
  domain. Approximation error increases as $a_0$ increases, especially
  in regions away from zero.}
  \label{fig:lmPriorZIn}
\end{figure}

The two plots in Figures~\ref{fig:lmPriorZIn} highlight the accuracy of
the mixture approximation in the region where MCMC samples are abundant
and the loss of accuracy in the tail. The left panel demonstrates
excellent agreement near $a_0 = 0$, where the posterior is
concentrated and samples abundant. The right panel shows that errors 
accumulate as $a_0$ increases; because of the mixture-density tails, 
minor deviations in the tail of the sampled values are amplified drastically.

\begin{figure}[htbp]
  \includegraphics[width=0.48\textwidth]{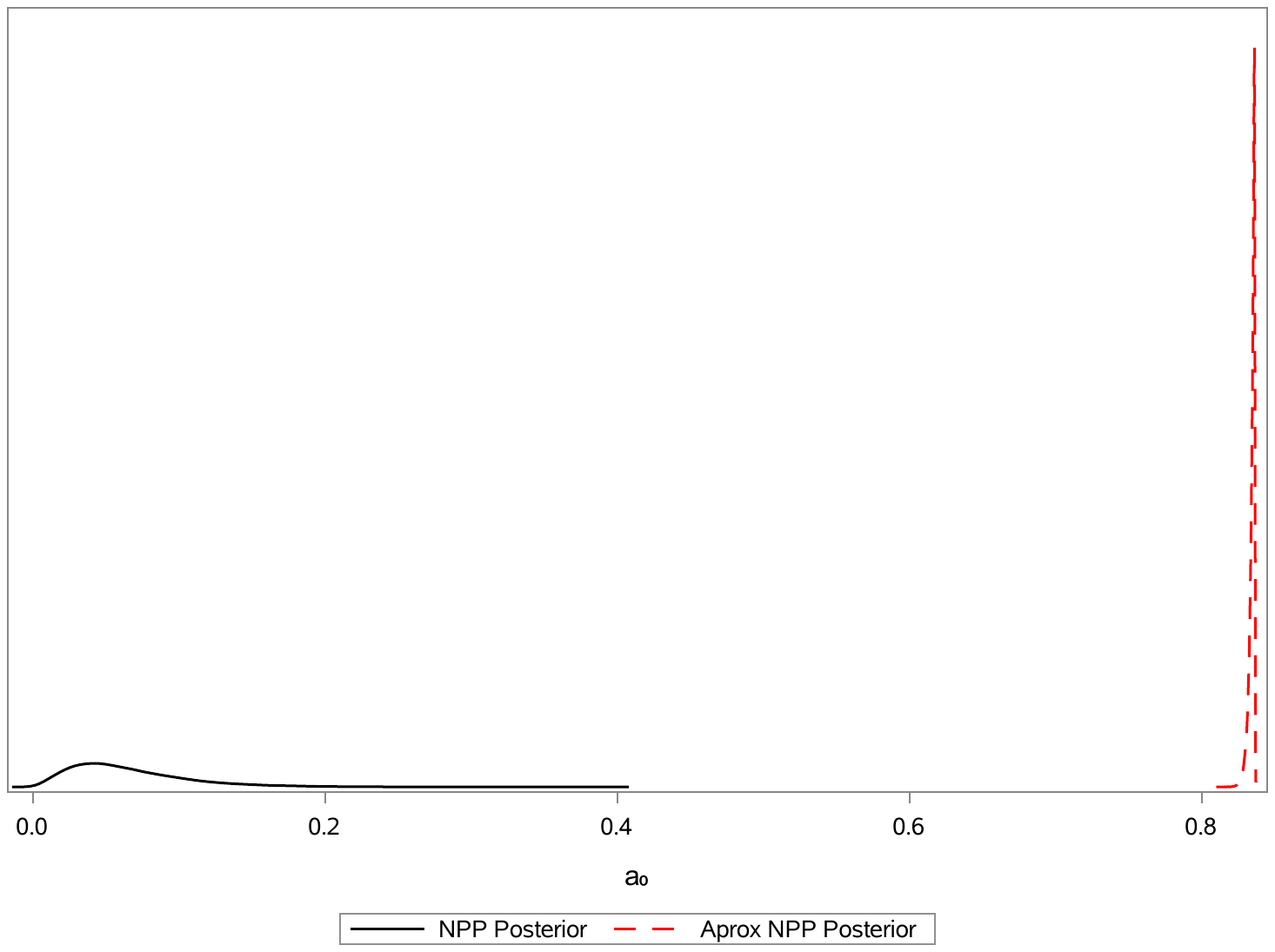}
  \hfill
  \includegraphics[width=0.48\textwidth]{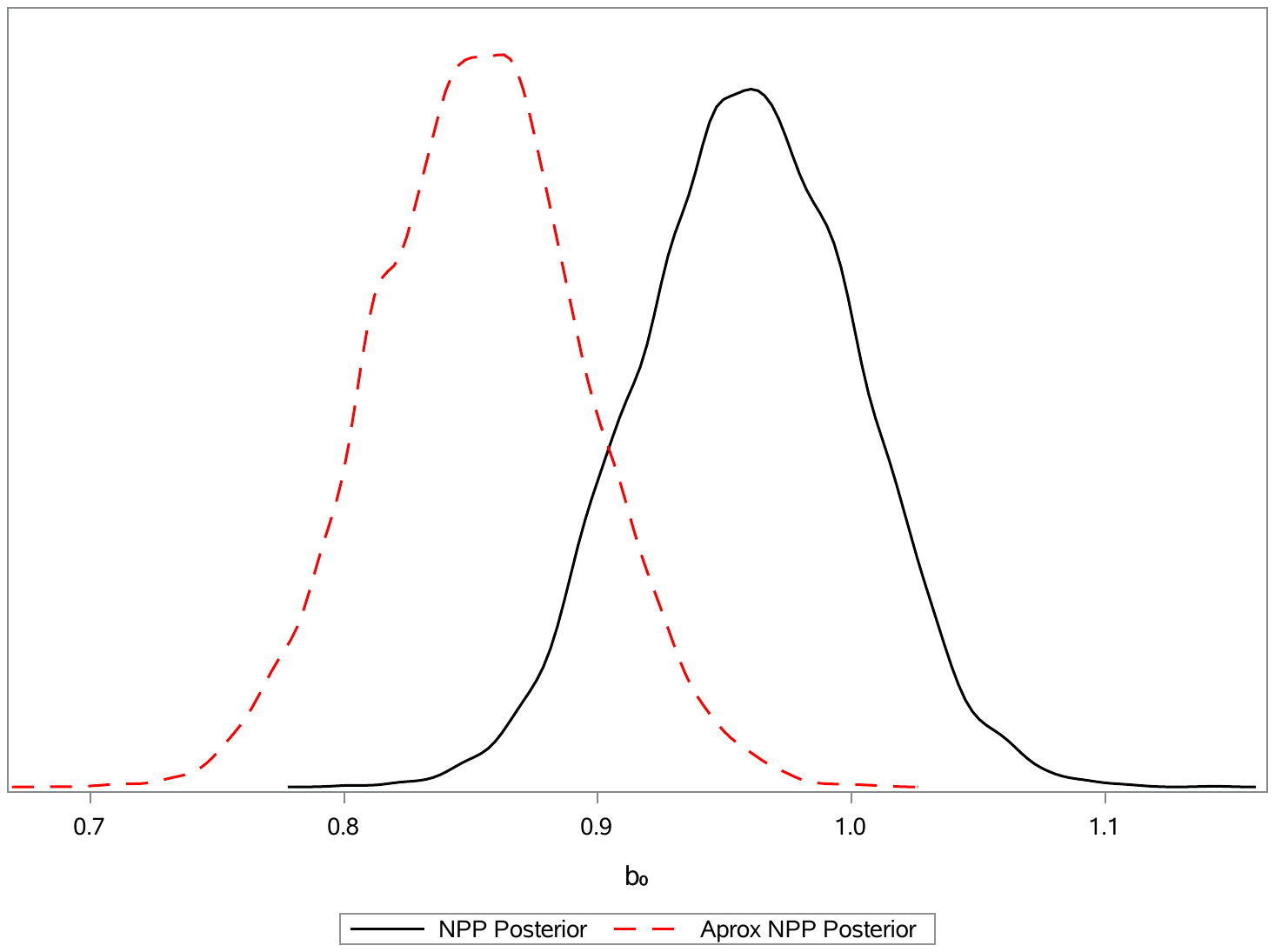}
  \caption{Left: Posterior distribution under a poorly estimated
  $C(a_0)$. Tail underestimation leads to spurious inflation of the
  posterior near large $a_0$ values. Right: incorrect posterior estimates
  on regression coefficients (only shown $b_0$ here).}
  \label{fig:lmWrongPost}
\end{figure}

Figure~\ref{fig:lmWrongPost} shows the consequence of a
wrongly-estimated $\log C(a_0)$. Because $C(a_0)$ appears in the
denominator of the NPP, an underestimated value artificially inflates
the posterior density and pushes $a_0$ to larger values. This leads to
over-borrowing, distorted posterior inference for $\theta$, and
misleading uncertainty quantification.

\subsection{Piecewise and Extrapolation Approaches}

This highlights a fundamental difficulty in sampling-based approximation 
methods such as the one we propose: one cannot reliably approximate a 
density in regions where no samples are observed. Because the logarithm 
density drops off so quickly, one can see that it would require an 
impractically large amount of computation time for a Markov chain to 
explore area where $a_0$ is large. When this
happens, though a mixture density can approximate the distribution well
where there are abundant samples, the tail approximation beyond certain
value is certainly to be off. The mixture tail can over- or
under-estimate the true function, with a very slim chance of getting it
right.

To correct the approximation in such cases, we exploit two structural
properties of the normalized power prior that provide anchor points.

\paragraph{Anchor at $a_0 = 0$.}
By definition,
\[
C(0)
=
\int \pi(\theta)\, d\theta
=
1,
\]
whenever the baseline prior $\pi(\theta)$ is proper.  Consequently,
\[
\log C(0) = 0,
\]
regardless of the historical data or the likelihood.  In practice, we
enforce this constraint by introducing a constant shift on the log scale
after fitting the mixture approximation.  All mixture-based evaluations of
$\log C(a_0)$ are re-centered to satisfy $\log \tilde C(0)=0$, ensuring
proper normalization of the power prior at the left boundary.

\paragraph{Anchor at $a_0 = 1$.}
At the other extreme,
\[
C(1)
=
\int L(\theta | D_0)\,\pi(\theta)\, d\theta,
\]
which is the marginal likelihood (normalizing constant) of an analysis
that uses only the historical data $D_0$ with baseline prior
$\pi(\theta)$.  Although this quantity is typically unavailable in closed
form, it can be approximated using, for example, the Laplace method
using MCMC samples. In this simulated regression example, the $\log(C(1)) = -367.539$
and the MCMC-sample-based estimate is $-367.473$. This approximation provides 
a stable and informative estimate of $\log C(1)$ that is independent of the UnPP sampling
behavior. 

The idea is to use a mixture to approximate the function in regions where samples are
abundant, and at a breakpoint $a_0^{*}$, use a linear extrapolation (on
the logarithm scale) to the log of the normalizing constant estimate of
$\log(C(1))$. The choice of $a_0^{*}$ is based on the sample, e.g. where
the mixture approximation starts to lose precision (see discussion in 
section \ref{sec:a0star}). In practice, you can
repeat UnPP MCMC simulations and mixture density estimation to see
where the curves start to diverge, as an indication to area where the
parametric approximation starts to lose precision. In practice, you can
use either the maximum observed value of $a_0$ in the UnPP MCMC samples
($a_0^{\max}$), or a fraction (e.g. 90\%) of that value. 

We define a piecewise approximation:
\[
\log \tilde C(a_0)
=
\begin{cases}
\log \tilde C_{\text{mix}}(a_0),
& a_0 \le a_0^{*}, \\[6pt]
m a_0 + b,
& a_0 > a_0^{*},
\end{cases}
\]
where the linear function is chosen to connect the two anchor points
\[
\bigl(a_0^{*}, \log \tilde C_{\text{mix}}(a_0^{*})\bigr)
\quad\text{and}\quad
\bigl(1, \log C(1)\bigr).
\]
The slope and intercept are given by
\[
m = \frac{\log C(1) - \log \tilde C_{\text{mix}}(a_0^{*})}{1 - a_0^{*}},
\qquad
b = \log C(1) - m.
\]

\begin{figure}[htbp]
  \includegraphics[width=0.48\textwidth]{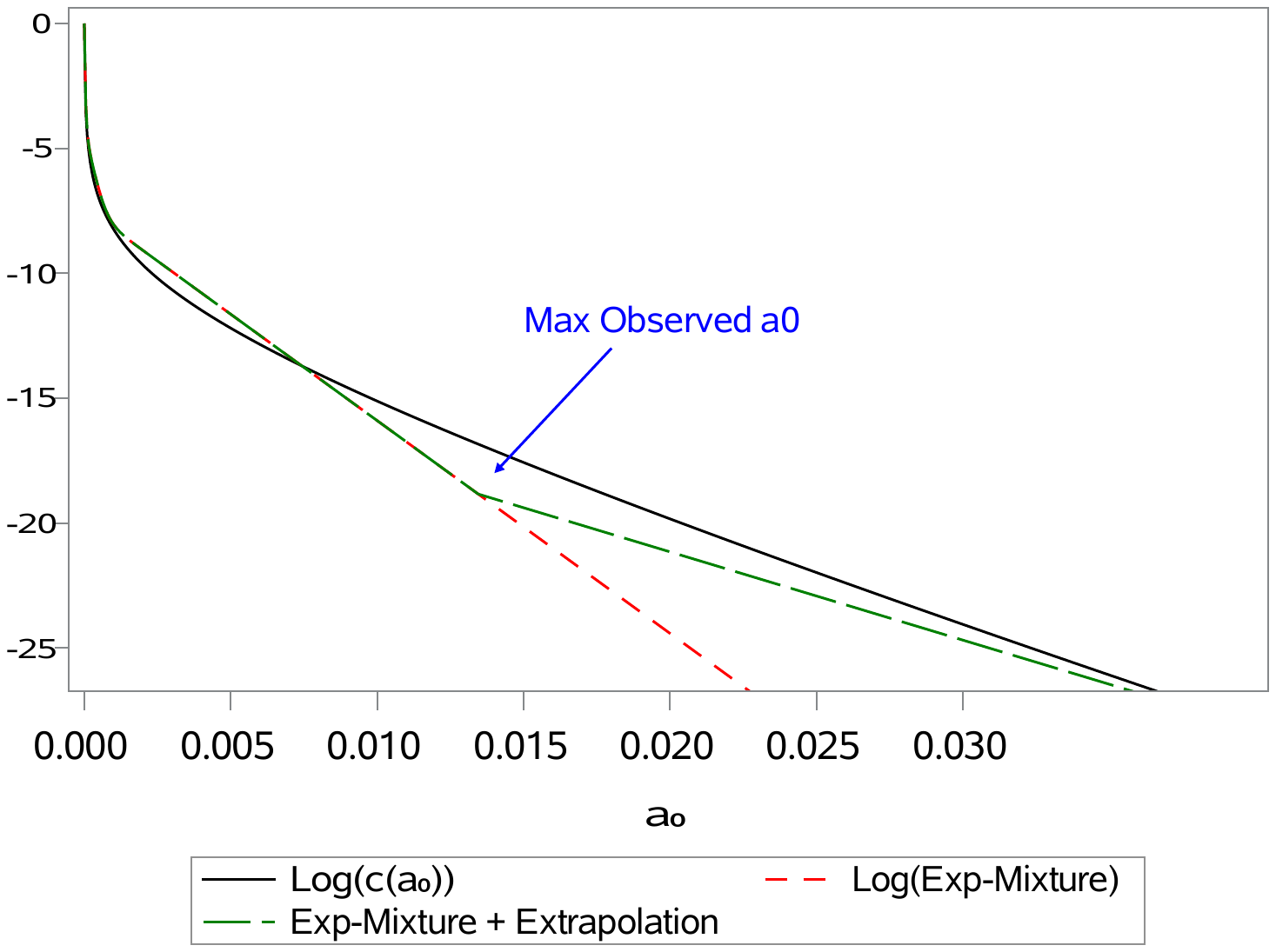}
\includegraphics[width=0.48\textwidth]{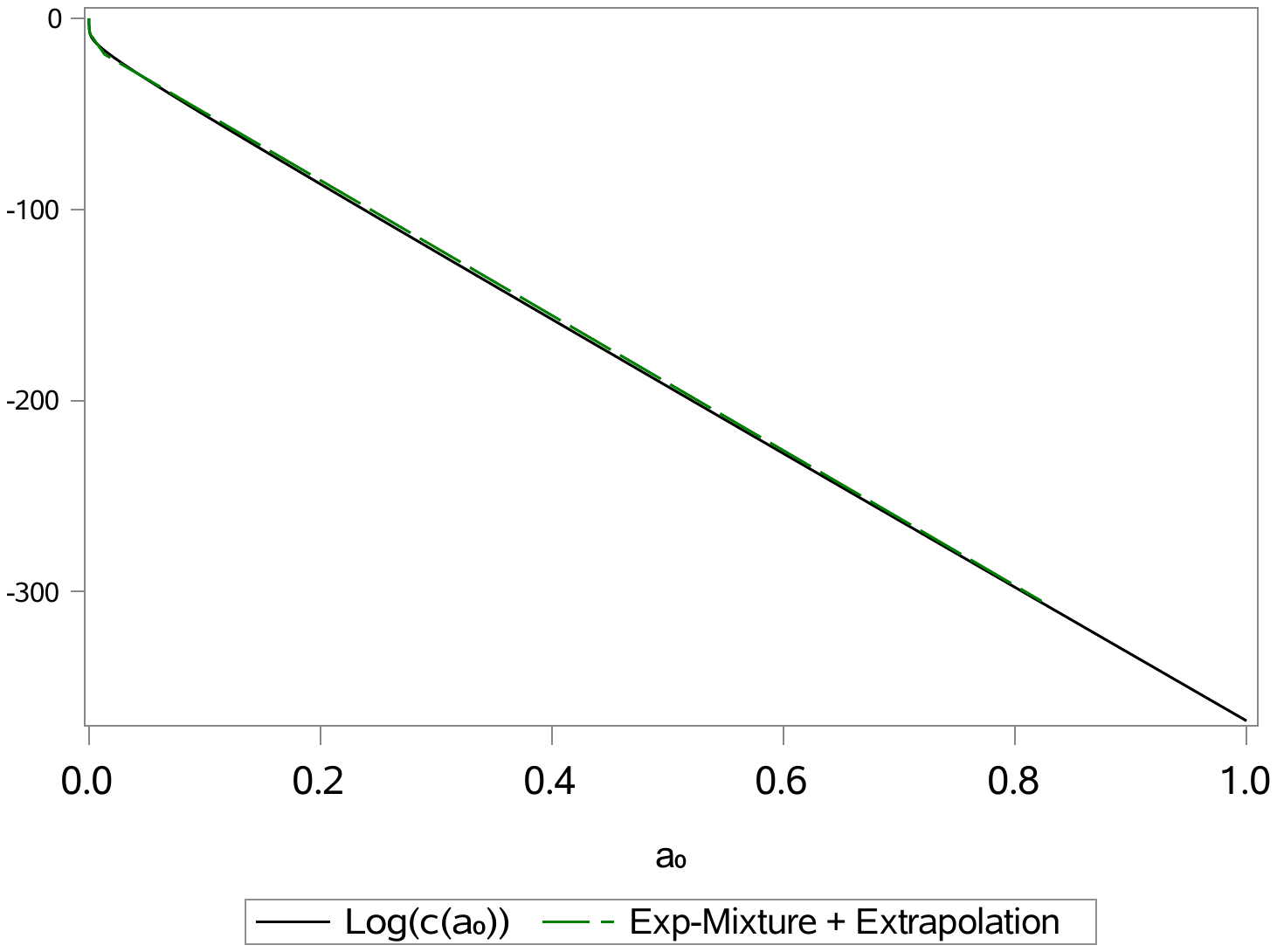}
  \caption{Left: at a selected point (here $a_0^{*} = a_0^{\max}$), we
  switch from a mixture approximation to linear extrapolation, to the
  logarithm of the normalizing constant at the full historical
  model. Right: zoomed out view of the true normalizing function and the
  mixture + extrapolation on the logarithm scale.}
  \label{fig:lmPriorZInAproxText}
\end{figure}

Figure \ref{fig:lmPriorZInAproxText} demonstrates how linear
extrapolation on the log scale works, where we choose $a_0^{*} =
a_0^{\max}$. The left panel shows the change point where we switch from
a mixture exponential distribution to an exponential tail (linear on the
logarithm scale). The right panel shows a zoomed-out view of the curve
over the entire range of 0 and 1. It is a significant improvement over
Figures~\ref{fig:lmPriorZIn}. The correction restores global agreement
and prevents spurious inflation of posterior mass at large
$a_0$ values.  

Taken together, this piecewise and extrapolation strategy leverages both
finite-sample information (from UnPP MCMC) and analytical structure
(from normalization at $a_0=0$ and Laplace approximation at $a_0=1$),
yielding a stable and practical approximation to $C(a_0)$.

\begin{figure}[htbp]
\centering
\includegraphics[width=4in]{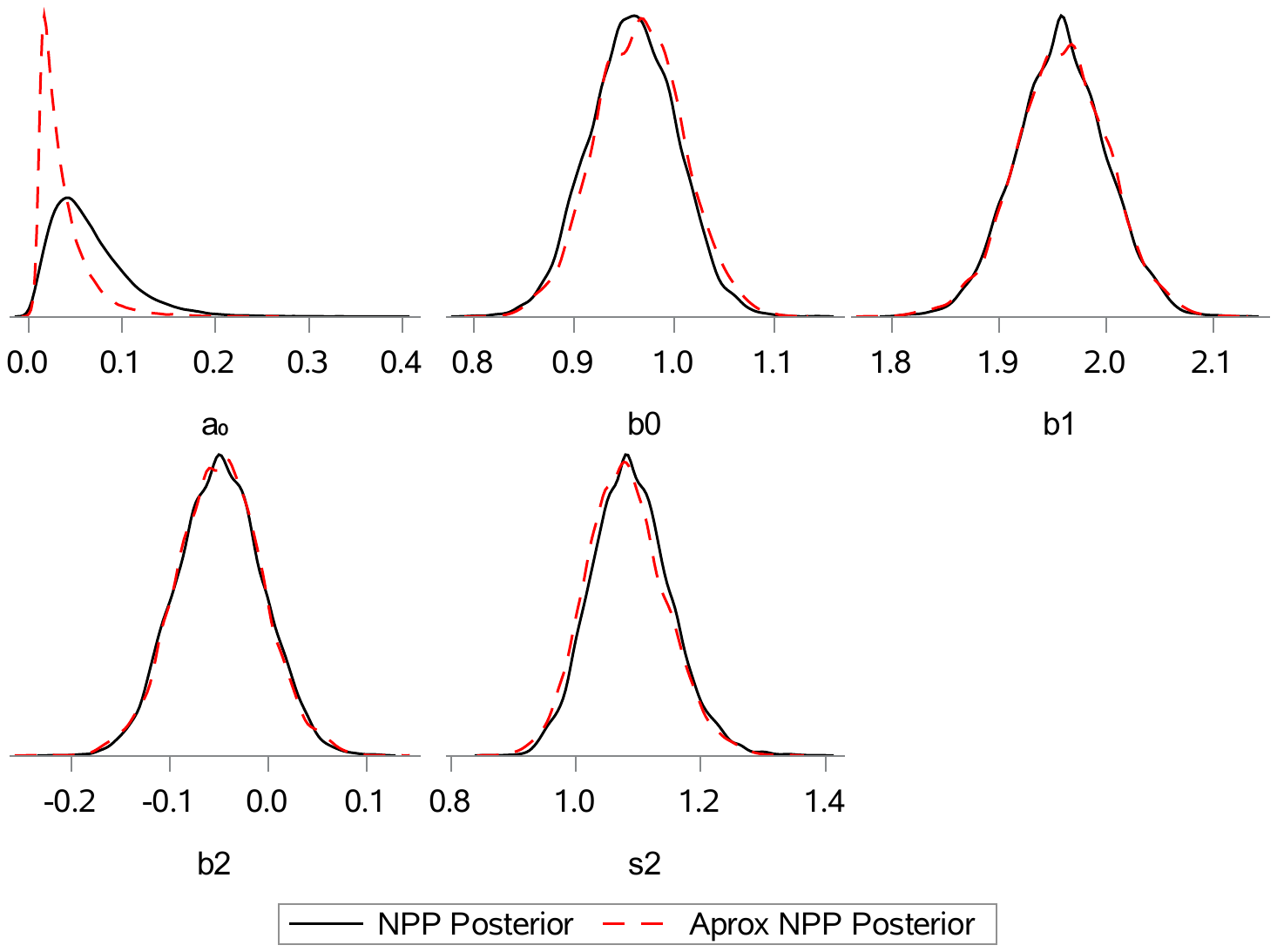}
  \caption{Posterior distribution under the true NPP (solid black curve)
  and its mixture-extrapolation-based approximation (dashed red
  curve). The posterior marginal distributions for $a_0$ are not
  precisely lined up, due to the approximation difference shown in
  Figure~\ref{fig:lmPriorZInAproxText}, the difference propagates
  little to the rest of the regression parameters as their posterior
  marginal distributions match up nicely.}
  \label{fig:lmPost}
\end{figure}

Figure~\ref{fig:lmPost} illustrates how combining mixture approximation
with extrapolation can remedy tail misspecification and improve
posterior estimates of parameters in the regression model. While the 
approximate density (red) differs from the true density (black), it 
overlaps strongly in the region where the true posterior has the highest 
probability. Because of this, the proposed method can lead to accurate 
posterior estimates on other model parameters and retain correct 
inferential properties. 

We know that $\log(C(a_0))$ is convex \citep{CarvalhoIbrahim2021},
therefore there is a limitation in a linear approximation, which can
only be precise up to a point, regardless how accurate one can compute
the normalizing constant at $a_0=1$ or select a good breakpoint in
$a_0^{*}$. However, this approximation method is computationally
convenient and executable in general Bayesian software. More
importantly, its impact on the model parameters is relatively mild and
substantially mitigates the distortions observed in
Figure~\ref{fig:lmWrongPost}.


\subsection{Choice of Breakpoint $a_0^{*}$}
\label{sec:a0star}

Selecting the breakpoint $a_0^{*}$ where the mixture approximation
transitions to linear extrapolation is a practical consideration. 
In principle, one could identify $a_0^{*}$ by running the UnPP program
multiple times, fitting mixture exponential distributions to each run,
plotting the fitted curves, and examining where the tails begin to
systematically deviate. This diagnostic approach would provide principled
evidence for choosing a breakpoint tailored to the specific data and
model.

However, this intensive procedure adds computational burden with minimal
benefit in practice. Because the true $C(a_0))$ is unavailable in practice, 
it is generally impossible to determine precisely where a parametric 
approximation begins to deviate. Modest misspecifications that relatively 
small region typically have negligible impact on inference for model parameters.

We prefer simplicity over exhaustive search. We suggest using either
$a_0^{*} = a_0^{\max}$ (the maximum observed value of $a_0$ in the UnPP
MCMC sample), or a fraction of that value, e.g. $a_0^{*} = 0.9 \times
a_0^{\max}$. This approach requires no additional fitting or
diagnostics, provides a reasonable anchor point for extrapolation.

\subsection{Sampling on a Transformed Scale}

When $C(a_0)$ decays sharply near zero, the UnPP posterior for $a_0$
concentrates in a narrow region, leaving the rest of $[0,1]$ sparsely
sampled and making mixture fitting numerically difficult. One remedy is
to sample on a transformed scale, e.g. log or logit. Consider the
reparameterization $\kappa_0 = \log(a_0)$, $a_0 = e^{\kappa_0}$ with
$\kappa_0 \in (-\infty, 0]$.

Define the transformed normalizing-constant function
\[
C_{\kappa}(\kappa_0)
=
\int L(\theta | D_0)^{e^{\kappa_0}} \pi_0(\theta)\,d\theta.
\]
Then the normalized power prior written on the $\kappa_0$ scale is
\[
\pi(\theta, \kappa_0 | D_0)
\propto
\frac{L(\theta | D_0)^{e^{\kappa_0}}}{C_{\kappa}(\kappa_0)}
\,\pi_0(\theta)\,\pi_A(e^{\kappa_0})\,e^{\kappa_0},
\]
where $e^{\kappa_0}$ is the Jacobian from $a_0=e^{\kappa_0}$.

For MCMC-based approximation on the transformed space, directly working
with $C_{\kappa}(\kappa_0)$ can be problematic because it is not
guaranteed to be integrable over $(-\infty,0]$. To obtain a proper
sampling target, we define the weighted function
\[
\bar C_{\kappa}(\kappa_0) = C_{\kappa}(\kappa_0)\,e^{\kappa_0}.
\]
We then approximate $\bar C_{\kappa}(\kappa_0)$, rather than
$C_{\kappa}(\kappa_0)$, from UnPP samples on the $\kappa_0$ scale.

To keep the resulting prior algebraically equivalent to the original
NPP, the inserted Jacobian factor must also appear in the numerator:
\[
\pi(\theta, \kappa_0 | D_0)
\propto
\frac{L(\theta | D_0)^{e^{\kappa_0}}\,e^{\kappa_0}}
  {\bar C_{\kappa}(\kappa_0)}
\,\pi_0(\theta)\,\pi_A(e^{\kappa_0})\,e^{\kappa_0}.
\]
Hence there are two Jacobian terms in the numerator with different
roles: one is the standard change-of-variable Jacobian; the other
cancels the Jacobian absorbed into $\bar C_{\kappa}(\kappa_0)$ for
computational tractability.

\subsection{Weighting the Normalizing Function}

This idea, adding a weight for computational convenience and then canceling it
in the prior specification, generalizes beyond the transformation
approach and motivates a broader weighting strategy. Let $w(a_0)>0$ be
any convenient weight on $[0,1]$, and define
\[
\bar C_w(a_0)=C(a_0)w(a_0).
\]
The key requirement for our MCMC-based approximation is that
$\bar C_w(a_0)$ be easier to sample and approximate than $C(a_0)$
itself. To preserve the original NPP formulation, the same weight must
be included in the prior numerator:
\[
\pi(\theta, a_0 | D_0)
\propto
\frac{L(\theta | D_0)^{a_0}w(a_0)}{\bar C_w(a_0)}
\,\pi_0(\theta)\,\pi_A(a_0),
\]
so the inserted weighting cancels algebraically.

One useful choice is to counter the rapid tail decay of $C(a_0)$ on the
log scale. Since $\log C(0)=0$ and $\log C(1)$ can be estimated from the
historical-only model, define the linear anchor
\[
\ell(a_0):=a_0\log C(1),
\]
and choose
\[
w(a_0):=\exp\{-\ell(a_0)\}=C(1)^{-a_0}.
\]
Then
\[
\log \bar C_w(a_0)=\log C(a_0)-a_0\log C(1),
\]
which standardizes the endpoint levels to
\[
\log \bar C_w(0)=0,\qquad \log \bar C_w(1)=0
\]
when $C(1)$ is exact, and approximately so when $C(1)$ is numerically
estimated. In practice, this reduces the log-scale dynamic range,
"lifts" the difficult tail region, and often produces a more manageable
shape for MCMC sampling and mixture approximation.

If additional stabilization is still needed, one can use an
annealing-like tempering step on the weighted target. Specifically,
define a tempered family
\[
\bar C_{w,\tau}(a_0) \propto \bar C_w(a_0)^{\tau}, \qquad 0<\tau\le 1.
\]
For $\tau<1$, the log-scale range is compressed,
\[
\log \bar C_{w,\tau}(a_0)=\tau\,\log \bar C_w(a_0),
\]
which lifts low-density regions and can substantially improve MCMC
exploration when the U-shape remains too deep after weighting. This
tempering should be viewed as a computational device for approximation,
not a change to the final inferential target: after fitting on the
tempered scale, one should map back to the original ($\tau=1$) target
through standard correction or by using tempering only to construct a
stable approximation. In practice, moderate values such as
$\tau\in[0.5,0.9]$ often provide a useful compromise between easier
sampling and approximation fidelity.

It is important to recognize that this remains a trade-off between
simplicity and accuracy. As we introduce more sophisticated weighting or
curve-shaping schemes to better approximate $C(a_0)$ (or a weighted
version of it), both computational cost and programming burden
increase. From a practical perspective, if the mixture-plus-
extrapolation approach already delivers robust and accurate-enough
inference, then it is often the most sensible and easiest strategy to
implement.

\section{Discussion}
\label{sec:discussion}

The normalized power prior (NPP) offers a principled Bayesian framework
for dynamically incorporating historical information while properly
accounting for uncertainty in the borrowing parameter $a_0$.  Despite
its theoretical appeal, the practical use of the NPP has been severely
limited in routine statistical analyses, primarily because the
normalizing constant $C(a_0)$ lacks a closed form in most models and is
computationally expensive to evaluate.  Existing methods that utilize
bridge and path sampling require external processing steps and are more
applicable to domain-specific software.  As a result, many practitioners
avoid using the normalized power prior. 

In this paper, we aimed to make the NPP fully compatible with
general-purpose Bayesian software such as SAS PROC MCMC, Stan, JAGS,
BUGS, or NIMBLE.  The central insight is that the UnPP posterior for
$a_0$ (based on historical data) is proportional to $C(a_0)$ when the
initial prior on $a_0$ is uniform. This enables a direct sampling-based
approximation of the normalizing constant function.  Once the shape of
$C(a_0)$ is captured through MCMC draws, it can be encoded as a
mixture distribution that is straightforward to evaluate inside a
standard Bayesian software package. In the common case where a mixture
fit from UnPP samples is adequate over $[0,1]$, the added computational
burden is relatively minor: fit a mixture model (e.g., via PROC FMM)
and insert $\log \tilde C(a_0)$ into an existing OPP Bayesian
program. This represents a practical trade-off, because without approximating $C(a_0)$ one typically cannot fit the NPP directly in general Bayesian
software. In more challenging cases, such as sharp decay near
$a_0=0$, additional steps are required (e.g., endpoint anchoring and
mixture-plus-extrapolation), but these steps are still feasible within
standard workflows.

The proposed method works well if the MCMC samples of $a_0$ from an
UnPP analysis of the historical data traverse the (0, 1) range well. In
that case, a mixture distribution can approximate the normalizing
constant function to a high degree of accuracy, and one can
replicate the NPP posterior with a simple added mixture function
evaluation in the software.

In other situations where the marginal posterior of $a_0$ in UnPP drops
off steeply near zero, simple sampling-based approximation may be
insufficient because the mixture tail may over- or under-estimate the
true $C(a_0)$. We therefore augment the mixture approximation with
extrapolated tail corrections derived from Laplace approximations at
$a_0=1$, beyond the range of observed samples. While this hybrid correction
carries additional computation relative to the base mixture approach, it remains
implementable in the same software environment and mitigates the primary
source of discrepancy between approximate and true $C(a_0)$, in area of moderate-to-large $a_0$ values that is weakly sampled by MCMC.

Three practical extensions are worth highlighting as potential
discussion points. First, breakpoint selection for piecewise
extrapolation can be tuned (e.g., using $a_0^{\max}$ or a fraction such
as $0.9\,a_0^{\max}$), but extensive search for an "optimal" breakpoint
often adds little value relative to its computational cost. Second,
sampling on a transformed scale (for example, $\kappa_0=\log(a_0)$) can
stabilize approximation when $C(a_0)$ is highly skewed, provided
Jacobian terms are handled consistently in the prior specification.
Third, weighting strategies, including annealing-like tempering of the
weighted target, can reduce dynamic range and improve MCMC exploration
in low-density regions. These refinements can improve robustness in hard
cases, but they also increase programming and computation. In practice,
the simplicity-accuracy trade-off remains central: if the
mixture-plus-extrapolation approach already provides stable and
accurate-enough inference, it is usually the most practical choice.

We showed through simulation that, while the mixture and extrapolation
approach provides an approximation to the true normalizing function, the
impact on the model parameters of interest is relatively little. This
work provides a practical and implementable computational strategy for
normalized power priors.  By leveraging MCMC sampling under the UnPP and
mixture-based density approximation, the NPP becomes accessible to
analysts using general Bayesian tools while retaining the flexibility
that makes it a valuable framework for dynamic borrowing.

\bibliographystyle{apalike}
\bibliography{powerprior}

\appendix 
\section{Appendix A: SAS Implementation}
\label{sec:appendix_sas}

This appendix provides the essential SAS code used to implement the original power prior (OPP), the unnormalized power prior (UnPP) (which is used to estimate the normalizing constant $C(a_0)$ via an exponential mixture), and the normalized power prior
(NPP) for both binomial and regression settings.  

\subsection*{Original Power Prior}

To implement the OPP, you combine the two data sets and carry out a regular Bayesian analysis. We assume the following are the historical and current data sets, and {\tt{alldata}} combines the two data sets. 

\begin{lstlisting}[language=SAS,basicstyle=\ttfamily\small]
data pilot;
   input event n;
datalines;
   20 100
;

data trials;
   input event n;
datalines;
   20 200
;

data alldata;
   set trials(in=i) pilot;
   if i then group="current";
   else group="pilot";
run;
\end{lstlisting}

The following code fits an OPP with a fixed $a_0$ (assume $a_0 = 0.3$) using the combined data set:

\begin{lstlisting}[language=SAS,basicstyle=\ttfamily\small]
proc mcmc data=alldata seed=17 nmc=500000 thin=10 
     outpost=oppOut;
   parm p 0.2;
   prior p ~ uniform(0, 1);
   a0 = 0.3;
   llike = logpdf("binomial", event, p, n);
   if (group = "pilot") then do;
      llike = a0 * llike;
   end;
   model general(llike);
   run;
\end{lstlisting}

\subsection*{Binomial Data}

Here we use the binomial data to estimate $C_0$. The UnPP uses only the historical data and places a Uniform $(0,1)$ prior on $a_0$. 

\begin{lstlisting}[language=SAS,basicstyle=\ttfamily\small]
proc mcmc data=pilot seed=17 nmc=500000 thin=10 
     outpost=histOut;
   parm p 0.2;
   parm a0 0.2;
   prior p a0 ~ uniform(0, 1);
   llike = a0 * logpdf("binomial", event, p, n);
   model general(llike);
   run;
\end{lstlisting}

This posterior sample of \texttt{a0} is proportional to the normalizing
function $C(a_0)$ and is used to construct its mixture approximation
(using \texttt{PROC FMM}, twice, once to find the optimal number of
components, once to obtain estimates given that component number). The
selection criterion for the optimal number of mixture components is by
BIC. Because $a_0$ is restricted to $[0,1]$, we use a bounded
exponential-mixture kernel on this interval in the NPP program.

\begin{lstlisting}[language=SAS,basicstyle=\ttfamily\small]
proc fmm data=histOut crit=BIC;
   model a0 = / kmax=6 dist=expo;
run;

proc fmm data=histOut crit=BIC;
   model a0 = / k=2 dist=expo; 
run;
\end{lstlisting}

which produces the following estimates:

\begin{lstlisting}[language=SAS,basicstyle=\ttfamily\small]
      Prob       Component    Estimate     scale

    0.1475               1     -3.6763    0.02532
    0.8525               2     -1.4380    0.23741
\end{lstlisting}

Once you obtain the mixture density estimates, fitting a NPP becomes
routine:

\begin{lstlisting}[language=SAS,basicstyle=\ttfamily\small]
proc mcmc data=allData outpost=AproxOut seed=15179 
     nmc=20000;
   parm p 0.2;
   prior p ~ uniform(0, 1);
   parm a0 0.2 / slice;
   beginprior;
   logC = log(  0.1475 * pdf("expon", a0, 0.02532)
              + 0.8525 * pdf("expon", a0, 0.23741));
   hyper a0 ~ general(-logC, lower=0, upper=1);
   endprior;
   llike = logpdf("binomial", event, p, n);
   if (group = "pilot") then do;
      llike = a0 * llike;
      end;
   model general(llike);
   run;
\end{lstlisting}

The difference between the NPP analysis and the OPP analysis presented in the first subsection is minimal. You treat $a_0$ as a parameter, assign an initial uniform prior, subtract the logarithm of the approximating normalizing constant, and the rest of the code remains the same. 

\subsection*{Mixture-Extrapolation Method}

In the situation where the extrapolation method is required, you need to
first compute the normalizing constant of a historical data model with
$a_0 = 1$. We approximate the posterior distribution using the multivariate normal with mean and covariance estimates, $\hat{\theta}$ and $\tilde{\Sigma}$, respectively. And the logarithm of the normalizing constant is the following:

\begin{align*}
q(\theta) &= \mathcal{N}(\theta | \mu = \hat{\theta}, \Sigma = \tilde{\Sigma}), \\
\log Z &= \log p(\mathbf{y}, \hat{\theta} | \mathbf{x}) + \frac{1}{2} \log |\tilde{\Sigma}| + \frac{D}{2} \log(2\pi),
\end{align*}
where $D$ is the dimension of the model. 

We use a regression example here for illustrative
purposes. The following are two simulated data sets, one for hist and one for current:

\begin{lstlisting}[language=SAS,basicstyle=\ttfamily\small]
data hist;
   call streaminit(3151);
   do j = 1 to 200;
      x0 = 1;
      x1 = rand("normal");
      x2 = rand("normal");
      mu = 0.5 + 2 * x1; * - x2;
      y = rand("normal", mu, sqrt(2));
      data = "hist";
      output;
      end;
   drop j mu;
run;

data curr;
   call streaminit(3114567);
   do j = 1 to 600;
      x0 = 1;
      x1 = rand("normal");
      x2 = rand("normal");
      mu = 1 + 2 * x1 - 0.7 * x2;
      y = rand("normal", mu, 1;
      data = "curr";
      output;
      end;
   drop j mu;
run;

data allData;
   set hist curr;
   run;
\end{lstlisting}

To compute the normalizing constant, you fit a linear regression using the {\tt{hist}} data set: 

\begin{lstlisting}[language=SAS,basicstyle=\ttfamily\small]
proc mcmc data=hist outpost=histOnly seed=15179 
     nmc=50000;
   parm b0 b1 b2;
   parms s2;
   prior b: ~ n(0, var=s2 * 1000);
   prior s2 ~ igamma(shape=2, scale=2);
   mu = b0 + b1 * x1 + b2 * x2;
   model y ~ normal(mu, var=s2); 
   run;
\end{lstlisting}

You compute the $\max(\mbox{\tt{logpost}})$ value from the MCMC
samples and the posterior covariance matrix, which you can get from the
autocall {\texttt{\%POSTCOV}} macro. The logarithm of the normalizing
constant is saved to the \texttt{logZ} macro. 

\begin{lstlisting}[language=SAS,basicstyle=\ttfamily\small]
proc sql noprint;
    select max(logpost) into :mlp_hist
    from histOnly;
quit;
%postcov(data=histOnly, var=b: s2, out=histcov);
%let np = 4;
proc iml;
   use histcov;
   read all var _num_ into cov;
   close histcov; 
   
   start logdet(A);
   G = root(A);
   return( 2*sum(log(vecdiag(G))) );
   finish;
       
   ldS = logdet(cov);
   lZ = &mlp_hist + ldS/2 + &np*log(2*constant("pi"))/2;
   call symputx("logZ", lZ);  
   quit;
%put &logZ;   
\end{lstlisting}

And the linear extrapolation equation coefficient $m$ and $b$ are
computed here (given the pre-specified $a_0^{*}$):

\begin{lstlisting}[language=SAS,basicstyle=\ttfamily\small]
%let m = %sysevalf((&logZ - &ldena0star)/(1 - &a0star));
%let b = %sysevalf(&logZ - &m );
\end{lstlisting}
where  {\texttt{\&ldena0star}} is the density evaluation of the
approximating mixture at {\texttt{\&a0star}}. 

The code needed to implement the NPP prior with extrapolation is listed
as the following: 

\begin{lstlisting}[language=SAS,basicstyle=\ttfamily\small]
   if a0 < &a0star then do;
      /* four-component mix-expon fit from PROC FMM:*/
      logC = log( p1*pdf("expon", a0, scale1) 
                + p2*pdf("expon", a0, scale2) 
                + p3*pdf("expon", a0, scale3) 
                + p4*pdf("expon", a0, scale4) );
      end;
   else do;
      logC = &m * a0 + &b;
      end;
   prior a0 ~ general(-logC, lower=0, upper=1);
\end{lstlisting}

\end{document}